# Integrated Surface-enhanced Raman Spectroscopy chip based on total-reflection liquid-core waveguide

Chunhong Lai[1,2], Li Chen[1], Junhui Li[1], Qinghao Liu[1], Shoumin Fang[3], Gang Chen[1,a]

[1] Key Laboratory of Optoelectronic Technology & Systems (Ministry of Education), Chongqing University, Chongqing 400044, China

[2] School of Electronic Information Engineering, China West Normal University, NanChong 637002, China

[3] School of Life Science, China West Normal University, NanChong 637002, China

[a] Author to whom correspondence should be sent. E-mail: gchen1@cqu.edu.cn

We propose a surface-enhanced-Raman-scattering(SERS) chip integrated with liquid-core waveguide. Utilizing the total-reflection of the liquid-core waveguide, it suppresses the leaky waveguide mode and enables a long propagation distance for SERS signal collection. Besides, the long SERS interaction distance has an average effect on the SERS signal along the waveguide, which improves the repeativity of the SERS measurement. 10 nM rhodamine 6G(R6G) Raman signal was obtain by QE65000 portable micro-spectrometer. An Raman enhancement factor for R6G of $2.5 \times 10^5$ was obtained with an excellent repeatability of 0.96% RSD (relative standard deviation). The SERS spectrum of silkworm DNA solution was also detected with our SERS chip, and the structural information was clearly shown. This chip provided a reliable testing way for studying DNA photosensitization structure damage.

Biochemical detection based on Raman spectroscopy has drawn considerable attentions for diverse applications in cancer diagnosis[1], drug analysis[2], explosives detection, water monitoring, and DNA detection[3]. When a laser interacts with a molecular, the scattering light contains the vibrational information via various shifted wavelengths, therefore, multi-component detection is possible based on the Raman spectrum. However, the detection sensitivity of conventional Raman spectroscopy is severely limited by its relative low Raman scattering cross section. Surface-enhanced Raman scattering (SERS) based on metallic nanoparticles[4] has been extensively studied for molecule identification at low concentration in micro-solution. Although several structures based on SERS have been reported[5], the trend to Lab on a Chip technology makes them unsuitable for integrating the SERS platform[6]. Besides, a further improvement of detection sensitivity and high repeatability of SERS chip is necessary for actual applications.

Recently, microfluidic devices based on waveguides have been utilized for SERS detection[7]. This technique is attractive in the sense that the laser propagates in the waveguides along with the sample molecule[8]. The multi-reflection provides a long effective interacting length to enhance the SERS signal. Several configurations have been reported based on photonic crystal fibers[9]. H Du et.al have demonstrated the detection of rhodamine 6G (R6G) with 100 pM[10] with photonic crystal fiber. However, it requires complex processing to manipulate in cladding holes, and it is hard to be integrated on a single chip. P Measor et.al presented a liquid-core anti-resonance reflecting optical waveguide based on $SiN/SiO_2$ layers, and a detection sensitivity of 30 nM has been achieved for R6G with silver nanoparticles solution[11]. However, this structure does not support total reflection, and the large propagation loss make the interacting length of this structure smaller than 0.7 mm, too short for sensitivity enhancement. To reach a relative high sensitivity, a high laser power was required, but this may cause damage of biomolecule. Recently we have demonstrated a 2 mm Si-based micro channel with inner wall coated with gold film, and a sensitivity of 10 nM for R6G was obtained[12]. However, the loss is still too large due to the leaky modes.

In this Letter, we demonstrated a SERS chip based on total-reflection liquid-core waveguide with large propagation distance but little loss. It allows a long interacting length and therefore accumulates the SERS signal, and the long interaction length also improve the repeatability of the Raman signal by averaging the SERS signal along the waveguide. The integration of the SERS substrate, micro-fluid, and liquid-core waveguide make such SERS chip a promising candidate for biochemical detection with high performance.

Figure 1(a) gives the structure of the proposed SERS chip. It comprises of a liquid-core waveguide, the sample inlet and outlet. The cross section of the liquid-core waveguide is depicted in Fig.1(b). The outer layer of the waveguide is a cylindrical micro-channel made by PDMS with a diameter of 600 μm. In order to achieve the total-reflection condition, the inner wall of the micro-channel was coated with a low refractive index material to form the total-reflection when the channel filled with water solution. The materials is AF 2400 with a refractive index of 1.29, which is smaller than the refractive index of water. Silver nanoparticles was fabricated by silver salt ($AgNO_3$) reduction with sodium citrate[13]. After centrifugation, the nanoparticles solution was injected into the waveguide, and was dried with a heating procedure, then it was coated on the inner wall of the waveguide. The excitation laser beam is injected into the waveguide from one end, and the final SERS signal is collected at the other end of the waveguide.

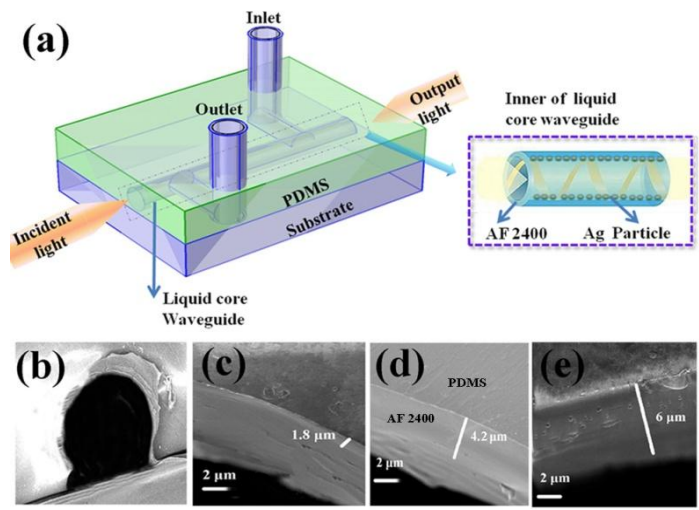

Fig. 1(a) Schematic of the SERS chip.(b) Waveguide cross section. (c)-(e) SEM of the cross sections with once, twice, and triple times coating of AF2400.

For the liquid core waveguide, the propagation loss must be depressed to ensure sufficient long interacting distance. Therefore, it is necessary to fabricate a waveguide supporting transmission mode while avoiding leaky mode. According to the calculation, in order to realize total reflection, the minimum thickness of AF 2400 is 5 μm[14]. Here, we coat the inner wall of the micro-channel with AF 2400 with three times[15]. The SEMs of the cross section images in Figs.1 (c)-(e) reveal that the thickness is 1.8 μm, 4.2 μm, and 6 μm, for coating once, twice, and triple times, respectively. And there is a clear interface between the two materials, indicating that AF 2400 has been well coated on the inner wall of the micro-channel. A further increases in the AF 2400 thickness resulted in rough surface, which degraded the waveguide performance. In the following experiments, the AF 2400 coating thickness is 6 μm.

Figure 2(a) illustrates the experiment setup for this SERS chip based on liquid core waveguide. A 632.8 nm He-Ne laser was used as the SERS excitation source with an output power of 10 mW. After alaser clean-up filter (F1), the beam was coupled into SERS chip from one end of the waveguide through a micro-objective lens (O1) with a numerical aperture (N.A.) of 0.25. From the other end of the waveguide, the output Raman signal was collimated by the another micro-objective lens (O2). After transmission of a dichroic beam splitter, the pure Raman light was coupled into optical fiber with 600 μm core diameter via a micro-objective lens (O3). Then it was sent into an Ocean Optics QE65000 portable micro-spectrometer for which signal-to-noise ratio is 1000:1 and resolution is 0.48 nm. All experiments were carried out with an integration time of 30 s. As shown in the Fig. 2(b), such a coupling system makes the waveguide functioning with excellent transmission mode, and the exciting laser propagates under total refection. We notice a bright light path, as most of the light propagates within the waveguide, and only little of it is scattered due to the roughness of the inner cladding.

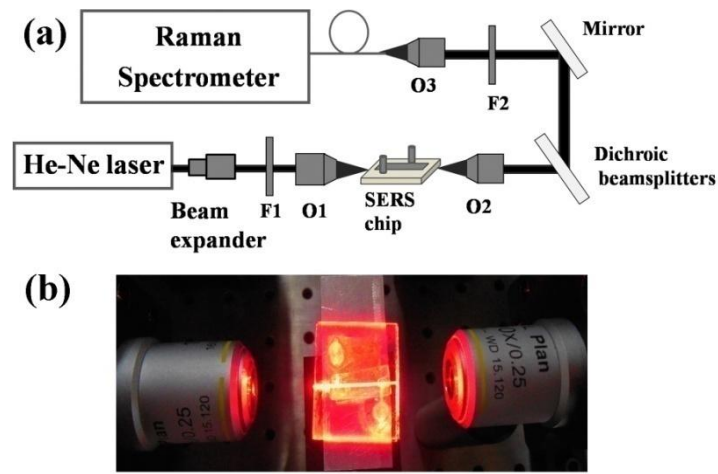

Fig. 2. (a) Schematic diagram of the detection setup (b) The chip under illumination

We used R6G to investigate the SERS chip performance. For comparison, experiments also carried out for the cases of pure R6G solution in nano-particle-coated waveguide and mixed solution of R6G and nano-particles in un-coated waveguide. It was found that the strongest SERS signal was obtained when mixed solution of R6G and nano-particles was applied in nano-particle-coated waveguide.

In order to study the relationship between the Raman enhanced performance and the waveguide length, for mixed R6G-nanoparticle solution ($10^{-6}$ M R6G), the Raman spectrum was obtained for different waveguide length. As shown in Fig. 3(a), the absolute intensity of the SERS signal of R6G at 1505 cm$^{-1}$ were 17346, 10329, 6100, 3347 at 1, 1.5, 2, 2.5 cm, respectively. The inset of figure 3(a) shown that the intensity with different length waveguide at 1505 cm$^{-1}$. When the waveguide length increased from 1 cm to 2.5 cm, Raman signal was reduced with the waveguide length, and the strongest SERS signal was obtained at a length of 1 cm. In the following of the paper, the waveguide length was 1 cm for all experiments, if not specified. In Fig. 3(b), the SERS signals were presented for mixed R6G-nanoparticle solution with various R6G concentrations. As shown in the figure, R6G solution can be detected at a concentration as low as 10 nM with this liquid-core waveguide based on SERS chip. It should be noted that, in this case, the length of the waveguide was about 1 cm, which was much larger than 2 mm in our previously result[12]. The detection limit of R6G was as low as 10 nM, and the intensity at 1505 cm$^{-1}$ was 690, which was about 8 times than our previously result[12]. It indicated that the SERS chip based on total-reflection can increase the propagation distance, improve the detection sensitivity.

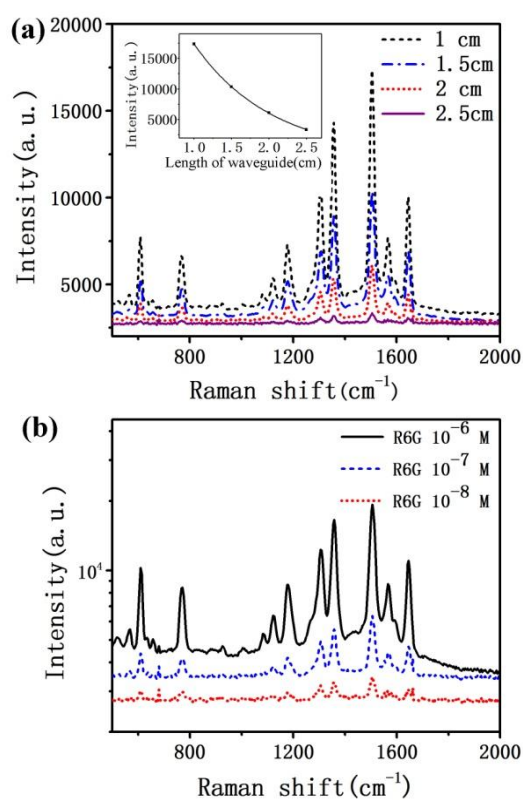

Fig. 3(a) SERS signals for different waveguide lengths (b) SERS signals for R6G with various concentration with 1 cm-long waveguide

The enhance factor (EF) is one of the key parameters in evaluating a SERS chip. In order to estimate the EF of this chip, firstly, a standard Raman spectrum of $10^{-2}$ M R6G solution (without mixed silver colloid) was taken with non-nanoparticle-coated waveguide. and then the SERS signal was obtained for the mixed R6G-nanoparticle solution($10^{-6}$ M R6G). The corresponding signals were plotted in figure 4. At the wavenumber of 1505 cm$^{-1}$, the peak absolute intensity of the SERS signal was 14566, while the standard Raman signal peak was 577. The enhancement factor can be calculated by EF=($I_{SERS}$/$C_{SERS}$)/($I_{RS}$/$C_{RS}$), where $I_{SERS}$ is the enhanced Raman signal, $C_{SERS}$ is the solution concentration for SERS, $I_{RS}$ is the standard Raman signal, and $C_{RS}$ is solution concentration of standard Raman. Therefore the enhancement factor is $2.5 \times 10^5$ for the SERS chip with 1 cm-long waveguide.

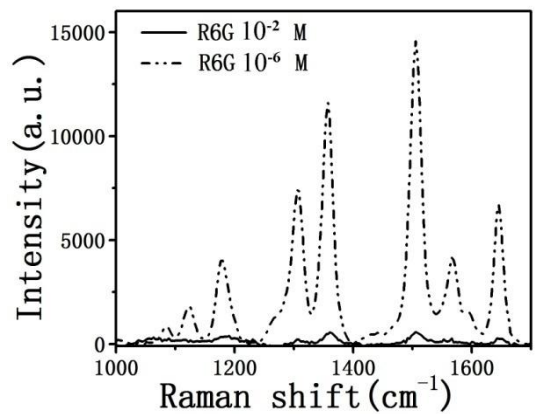

Fig.4 Enhanced Raman spectrum of $10^{-6}$ M rhodamine 6G mixture solution with SERS chip and Standard Raman spectrum of $10^{-2}$ M R6G with liquid-core waveguide (without silver nanoparticle)

The repeatability of the detection is vital for practical application of SERS chip. Because the chip with 1.5 cm-length was more robust and easy cleaning than that with 1.0 cm , here, we measure the Raman signal's repeatability with a 1.5cm-long chip. For the same concentration of $10^{-6}$ M R6G mixed with silver nanoparticles, experiments were done on the same SERS chip. Firstly, we have tested the SERS signal every 30 s, six results as show in Fig. 5 (a). The spectrum lines were nearly same. At 1505 cm$^{-1}$, the intensity values were 336, 339, 332, 335, 337, and 329. After simple calculation, the standard deviation and relative standard deviation are 3.2 and 0.96%. Then, after each test, the waveguide was cleaned with ethanol followed by deionized water washing to get rid of residual R6G, repeatedly tested six times. The standard deviation of six measurements was used as error bars, as shown in Fig. 5 (b). The standard deviation are 6 and 13 for peak at 1358 cm$^{-1}$ and 1505 cm$^{-1}$, respectively. The low standard deviation demonstrate that this SERS chip was repeatable and can be utilized for quantitative detection of analytes.

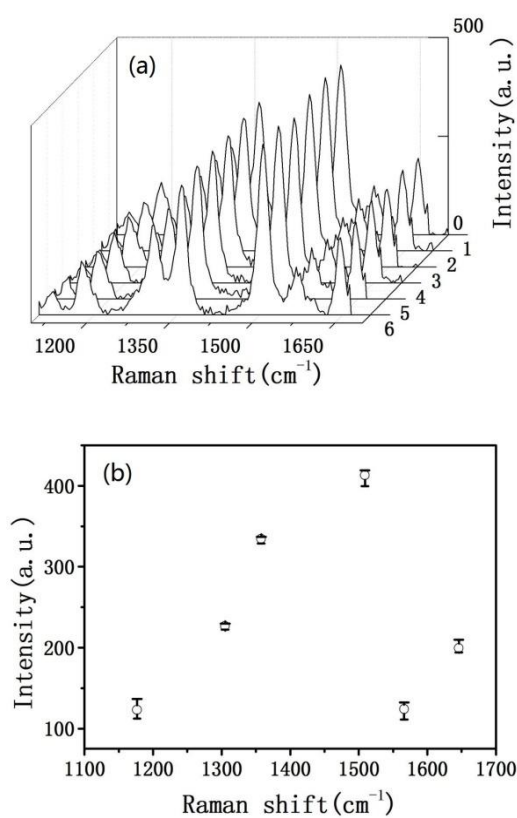

Fig. 5(a) SERS signals were tested six times by every 30s (b) SERS signal intensity for wavenumber at 1177,1305,1358,1505,1566 and 1645. Error bars represent standard deviation.

This integrated SERS chip is suitable for detecting bio-molecules in small quantity. Fig. 6 gives the SERS of 3 μL silkworm DNA obtained by SERS chip at a concentration of 25 μg/mL. The peaks were clearly shown at 226, 680, 813, 925, 1255, and 1566 cm$^{-1}$. The low wavenumber 226 was enhanced due to the chemical absorption between DNA samples and the silver nanoparticles. The peak 680 cm$^{-1}$ was from stretching vibration of base adenine A[16], and 813 cm$^{-1}$ was caused by the symmetrical stretching of phosphodiester (OPO), and 1566 cm$^{-1}$ came from the stretching vibration of the C=C bond. It clearly illustrated the fine information of the silkworm DNA molecule structure. This chip was suit for detecting biomolecule. It provided a reliable testing way for studying DNA photosensitization damage and microscopic damage characteristics of DNA in aqueous solution.

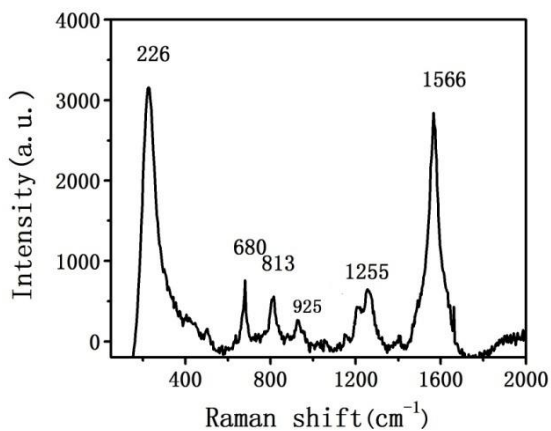

Fig. 6 Raman spectroscopy of silkworm DNA.

In conclusion, we have demonstrated a SERS chip based on total-reflection liquid-core waveguide, which can increase the propagation distance and improve the detection sensitivity. This chip achieved the detection limit of 10 nM R6G solution by QE65000 portable micro-spectrometer. And it has a good enhancement performance which the EF for R6G is $2.5 \times 10^5$. This integrated chip exhibits excellent repeatability of 0.96% RSD. Using this SERS chip, we successfully obtained the Raman spectrum of silkworm DNA molecule solution. However, the improvement of the detection performance is still needed, for example, to further decreas the propagation loss through optimizing the coating of AF 2400, and a self-assemble of the silver nanoparticles on the inner wall of the waveguide will also increase the sensitivity.

[1] W. B. Wang, J. H. Zhao, M. Short, and H. S. Zeng, J. Biophotonics .**8** (7), 527 (2015).

[2] T. Frosch and J. Popp, J.Biomed.Optics.**15** (4) (2010).

[3] J. Bansal, I. Singh, P. K. Bhatnagar, and P. C. Mathur, J. Biosci. Bioeng. **115** (4), 438 (2013).

[4] Martin Moskovits, Rev. Mod. Phys. **57** (3), 783 (1985).

[5] C.H. Lai, G. Chen, L. Chen, X.S.Zhang, H.X. Zhang, and Y. Xu, Spectrosc. Lett. **49** (1), 51 (2015).

[6] N. Qi, B. W. Li, H. Y. You, W. Zhang, L. W. Fu, Y. Q. Wang, and L. X. Chen, Anal. Methods **6** (12), 4077 (2014).

[7] C. C. Fu, Y. J. Gu, Z. Y. Wu, Y. Y. Wang, S. P. Xu, and W. Q. Xu, Sensors and Actuators B-Chemical **201**, 173 (2014).

[8] S. J. Pearce, M. E. Pollard, S. Z. Oo, R. Chen, and M. D. B. Charlton, Appl. Phys. Lett. **105** (18), 181101 (2014).

[9] Y. Han, S. L. Tan, M. K. K. Oo, D. Pristinski, S. Sukhishvili, and H. Du, Adv. Mater. **22** (24), 2647 (2010).

[10] Yun Han, Opt. Eng. **47** (4), 040502 (2008).

[11] P. Measor, L. Seballos, D.L. Yin, J. Z. Zhang, E. J. Lunt, A.R. Hawkins, and H. Schmidt, Appl. Phys. Lett. **90** (21), 211107 (2007).

[12] C. H. Lai, L. Chen, G. Chen, Y. Xu, and C. Y. Wang, Appl Spectroscopy **68** (1), 124 (2014).

[13] P. C. Lee and D. Meisel, J. Phys. Chem. **86** (17), 3391 (1982).

[14] P. Dress, M. Belz, K. F. Klein, K. T. V. Grattan, and H. Franke, Sensors and Actuators B-Chemical **51** (1-3), 278 (1998).

[15] H. Bissett and H. M. Krieg, S. Afr. J. Sci. **109**,9 (2013).

[16] G.J.Thpmas, and Y.Kyogoku, Pract. Spectrosc. Ser. **1C**,717(1977).